\documentclass[twocolumn,aps,prl,assymb,amsmath,longbibliography]{revtex4}

\usepackage{color}
\usepackage{amsmath}
\usepackage{amssymb}
\usepackage{graphicx}
\usepackage[colorlinks=true,linkcolor=red]{hyperref}
\usepackage[sort&compress]{natbib}
\usepackage[dvipsnames]{xcolor}
\usepackage{soul}
\usepackage{amsfonts}
\usepackage{float}
\usepackage{color}
\usepackage{ulem}

\begin{document}

\title{Topological flat Wannier-Stark bands}

\author{A.~R.~Kolovsky$^{1,2,3}$, A.~Ramachandran$^3$, and S.~Flach$^3$}
\affiliation{$^1$Kirensky Institute of Physics, 660036 Krasnoyarsk, Russia}
\affiliation{$^2$Siberian Federal University, 660041 Krasnoyarsk, Russia}
\affiliation{$^3$IBS Center for Theoretical Physics of Complex Systems, 34051 Daejeon, South Korea}
\date{\today}
\begin{abstract}
We analyze the spectrum and eigenstates of a quantum particle in a bipartite two-dimensional tight-binding dice network with short range hopping under the action of a dc bias. We find that the energy spectrum consists of a periodic repetition of one-dimensional energy band multiplets, with one member in the multiplet being strictly flat. The corresponding macroscopic degeneracy invokes eigenstates localized exponentially perpendicular to the dc field direction, and super-exponentially along the dc field direction. 
We also show that the band multiplet is characterized by a topological winding number (Zak phase), which changes abruptly if we vary the dc field strength. These changes are induced by gap closings between
the flat and dispersive bands, and reflect the number of these closings.
\end{abstract}
\maketitle

\begin{figure}[h]
\includegraphics[width=7.5cm]{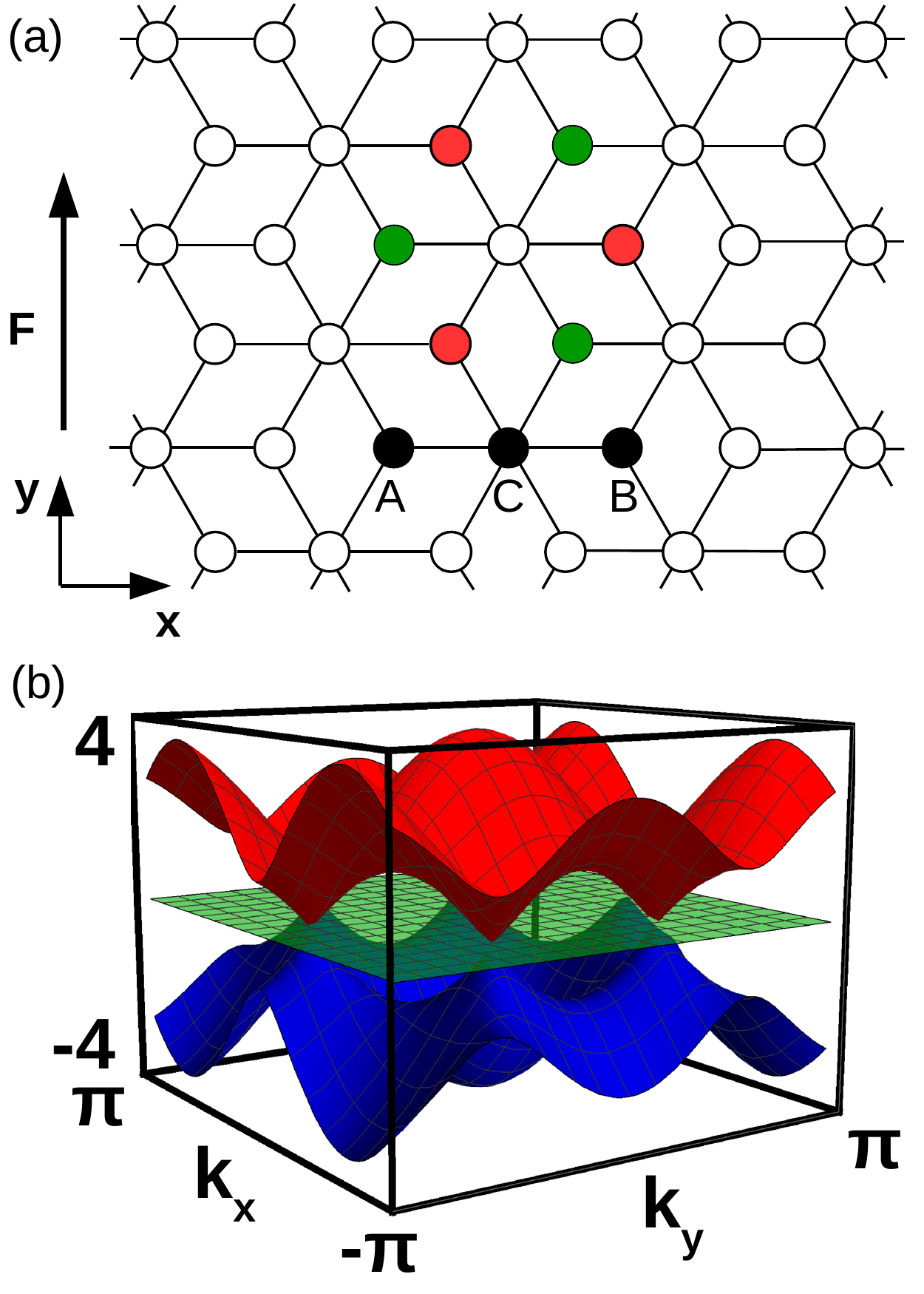}
\caption{(a) The dice lattice with the elementary cell consisting of three sites denoted by $A$, $B$, and $C$. The red and green circles represent the amplitudes $+ 1/\sqrt{6}$ and $- 1/\sqrt{6}$ respectively of the compact localized states. 
The black sites indicate an uncoupled trimer in the limit of infinitely strong dc bias.
(b) Bloch bands of the dice lattice with a flat band at $E=0$.}
\label{fig1}
\end{figure}

Recently much attention has been paid to flat bands in one-, two- and three-dimensional lattices with short range hoppings and non-trivial geometry~\cite{derzhko2015strongly}.  Flat bands with finite range hoppings exist due to destructive interference leading to a macroscopic number of degenerate compact localized eigenstates (CLS) which have strictly zero amplitudes outside a finite region of the lattice \cite{flach2014detangling}. Flat band networks have been proposed in one, two, and three dimensions and various flat band generators were identified~\cite{mielke1991ferromagnetism,tasaki1992ferromagnetism,maimaiti2017compact,Ramachandran:2017aa}.  Experimental observations of flat bands and CLS are reported in photonic waveguide networks~\cite{guzman2014experimental,vicencio2015observation,mukherjee2015observation,mukherjee2015observation1,Mukherjee:17,weimann2016transport,xia2016demonstration,Real:2017aa,travkin2017compact}, 
exciton-polariton condensates~\cite{masumoto2012exciton,baboux2016bosonic,whittaker2017exciton},
and ultracold atomic condensates~\cite{taie2015coherent,jo2012ultracold}.  
The tight binding network equations correspond to an eigenvalue problem $E\Psi_l = -\sum_m t_{lm}\Psi_m$. For bipartite lattices, 
the existence of flat bands and CLS is ensured through a proper usage of the protecting chiral symmetry \cite{Ramachandran:2017aa}. 
For example,   for the dice lattice shown in Fig.~\ref{fig1}(a) and $t_{lm}=1$ the CLS consists of an empty $C$ site which is surrounded by six excited $A$ and $B$ sites with alternating  amplitudes $\pm 1/\sqrt{6}$.

When a dc field is added, quantum particles start to experience well studied Bloch oscillations (see e.g.
\cite{kolovsky2004bloch,morsch2006dynamics}). Our results are therefore applicable for ultracold atoms in optical lattices where the electric field is
substituted by a tilt of the lattice in the gravitational field~\cite{anderson1998macroscopic} or accelerating the whole lattice~\cite{PhysRevA.65.063612}.
The same type of perturbations can be arranged in optical waveguide arrays where the electric field is modeled by a curved geometry of the waveguides~\cite{longhi2006observation}.

In the present letter we report on a new family of flat bands which exist in dc biased bipartite lattices and which are {\sl not} supported by CLS, despite of the short range hopping. We consider the dice lattice in the presence of a dc field oriented along the $y$-direction
(Fig.\ref{fig1}(a)). The absence of CLS indicates
nontrivial topology. We compute winding numbers (Zak phase) which indeed show abrupt changes through conical intersection point degeneracies upon variations of the dc field
strength.
We also generalize to 
different orientations of the field ${\bf F}$ and other lattice geometries.

\begin{figure}
\includegraphics[width=8.5cm]{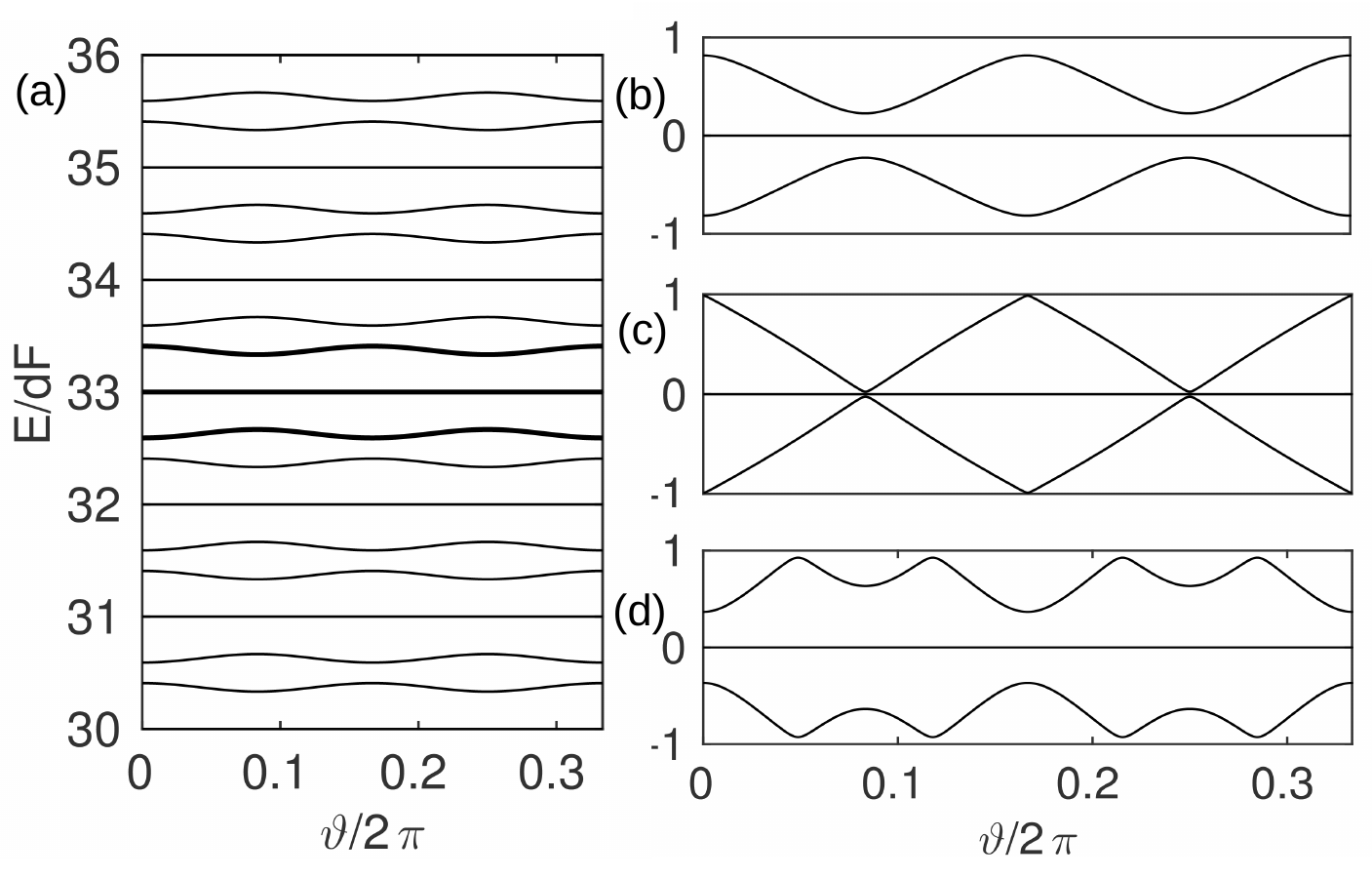}
\caption{ (a) Wannier-Stark band ladder of the biased dice lattice with ${\bf F}$ along the $y$ axis, $F=4$. Thick lines highlight the irreducible triplet of three Wannier-Stark bands. (b-d)  Irreducible  triplets for (b) $F=2$, (c) $F=1/0.62$, and  (d) $F=1$. }
\label{fig2}
\end{figure}

It is instructive to consider first the formal limit $|{\bf F}| = F\rightarrow\infty$. In this limit the particle energy is given by a Wannier-Stark ladder of triplets $(-\sqrt{2}t+ndF,ndF,\sqrt{2}t+ndF)$ where $t$ is the hopping matrix element and $d=\sqrt{3}/2$ is the Stark lattice period, i.e., the distance between rows of sites with the same Stark energy (here we set the fundamental lattice period to unity) and $n$ is the integer counting of the ladder steps. Furthermore, a quantum particle is confined to uncoupled trimers $A-C-B$ (shown with black circles in Fig. \ref{fig1}(a)). However, for a finite $F$ the particle can tunnel to neighboring raws. Recovering of tunneling together with translational symmetry allows to search for the particle eigenstates as Bloch waves  in the direction orthogonal to ${\bf F}$ \cite{doi:10.1143/JPSJ.62.2773,PhysRevA.87.033602}. This statement implicitly assumes `rational' orientations of the field, where ${\bf F}$ is parallel to a line connecting any two sites of the lattice separated by a finite distance \cite{PhysRevA.87.033602}. The energy spectrum then consists of a ladder of one-dimensional bands -- the Wannier-Stark bands. For the considered example these bands are shown in Fig.~\ref{fig2}(a) for $F=4$. It is seen in Fig.~\ref{fig2} that every third band is flat. In what follows we explain the presence of flat Wannier-Stark bands in the energy spectrum and obtain localized states associated with these bands. 

\begin{figure}[t]
\includegraphics[width=7.5cm]{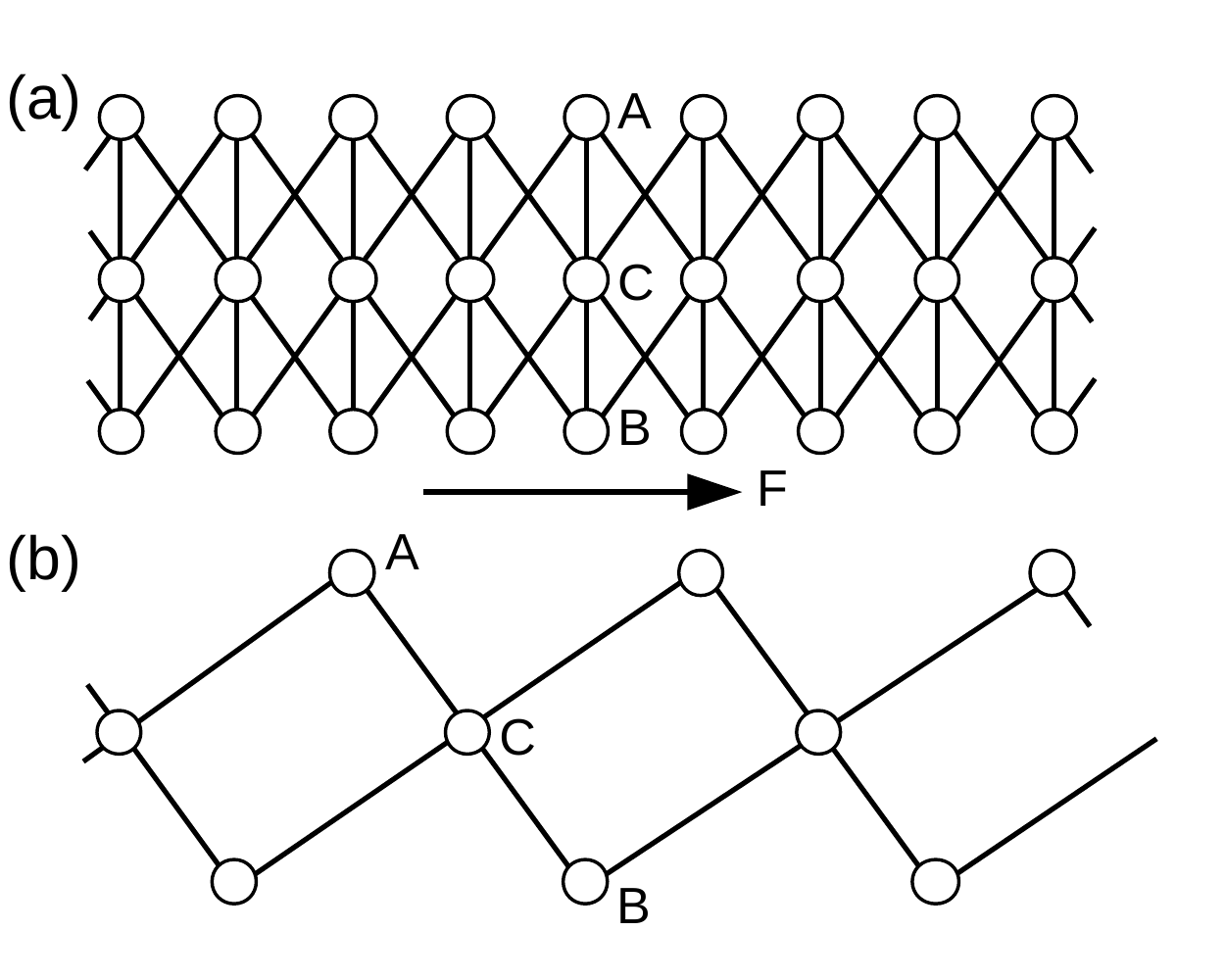}
\caption{The dual ladder system for the dice lattice: (a)  tilted in the $y$ direction, and (b) in the $x$ direction. }
\label{fig3}
\end{figure}

The stationary Schr\"odinger equation for the biased dice lattice reads
\begin{equation}
\label{b0}
E\Psi({\bf R}_j)=({\bf F}\cdot{\bf R}_j)\Psi({\bf R}_j)
-t\sum_{m}\Psi({\bf R}_j+{\bf r}_m)\;,
\end{equation}
where ${\bf R}_j$ are the lattice site positions and $|{\bf r}_m|=1$.  The sum over ${\bf r}_m$ connects three neighboring sites for $A$  or $B$ sites and six neighboring sites for $C$ sites (see Fig.\ref{fig1}).  Keeping in mind  that ${\bf F}$ is parallel to  the $y$ axis, we use the substitution $\Psi({\bf R}_j)\sim\exp(i\kappa R_j^{x})\psi(R_j^y)$ where $\kappa$ is the transverse quasimomentum. That reduces the original eigenvalue problem (\ref{b0}) to a one-dimensional eigenvalue problem with infinitely many bands: 
\begin{widetext}
\begin{eqnarray}
\nonumber
&E\psi^A_p&=dFp\psi^A_p - t\psi^C_p e^{-i2\vartheta} - t(\psi^C_{p+1}+\psi^C_{p-1}) e^{i\vartheta} \;, \\
\label{b1}
&E\psi^C_p&=dFp\psi^C_p - t\psi^A_p e^{i2\vartheta} - t \psi^B_p e^{-i2\vartheta} 
 - t(\psi^A_{p+1}+\psi^A_{p-1}) e^{-i\vartheta} - t(\psi^B_{p+1}+\psi^B_{p-1}) e^{i\vartheta} \;, \\
\nonumber
&E\psi^B_p&=dFp\psi^B_p - t\psi^C_p e^{i2\vartheta} - t(\psi^C_{p+1}+\psi^C_{p-1}) e^{-i\vartheta}  \;,
\end{eqnarray}
\end{widetext}
where  $\vartheta=a\kappa$ and $a=1/2$ is the distance between columns of sites. The system (\ref{b1}) is a three-leg ladder in a static field aligned with the ladder legs, see Fig.~\ref{fig3}.   This is a general result which holds for any rational orientation of the dc field and any $\nu$-band lattice in two dimensions:  in the Bloch representation for the wave vector orthogonal to the dc field, each Bloch  wave number yields a dual one-dimensional $\nu$-leg ladder in the direction of the dc field. The precise network topology of the dual ladder is depending on the field orientation, see Fig.\ref{fig3}.

The eigenvalue problem (\ref{b1}) for the dual system can be analyzed using the method of generating functions \cite{PhysRevA.91.053631,PhysRevA.91.053632}. We introduce the Fourier series expansion of three periodic functions in  $\theta$:
\begin{equation}
 Y^{A,B,C}(\theta)=(2\pi)^{-1/2} \sum_{p=-\infty}^\infty \psi_p^{A,B,C}\exp(ip\theta) \;,
\label{vectorfunction}
\end{equation}
and arrange them into a column vector function ${\bf Y}=[Y^A,Y^C,Y^B]^T$. Then the system of linear algebraic equations (\ref{b1}) is generated from the following differential equation set:
\begin{equation}
\label{b3}
idF\frac{{\rm d} {\bf Y}}{{\rm d}\theta}=G(\theta;\vartheta){\bf Y} \;,
\end{equation}
where
\begin{equation}
\label{b4}
G(\theta;\vartheta)=-\left(
\begin{array}{ccc}
E & f & 0 \\
f^* & E & f \\
0 & f^* & E
\end{array} \right)
\end{equation}
and $f(\theta;\vartheta)=t\exp(-i2\vartheta) + 2t\cos(\theta)\exp(i\vartheta)$. 
Consider now the unitary map ${\bf Y}(2\pi) = U(\vartheta,E) {\bf Y}(0)$ generated by (\ref{b3}). We are searching for an initial vector ${\bf Y}(0)$ and energy $E$ which satisfy periodicity ${\bf Y}(2\pi) = {\bf Y}(0)$. This can be achieved by first setting $E=0$, and computing the eigenvalues $\lambda_j$ and eigenvectors ${\bf Y}_{j}$ of $U(\vartheta,0)$ with $j=1,2,3$.  Together with the gauge invariance of (\ref{b3}) it follows that periodicity is obtained for initial conditions ${\bf Y}(0)={\bf Y}_j$ and energies 
\begin{equation}
\label{b7}
E_n^{(j)}(\vartheta)= dFn + i \frac{dF}{2\pi} \ln \lambda_j(\vartheta) \;,\quad -\infty<n<\infty \;.
\end{equation}

Next we note that for $E=0$ the matrix $G$ can be presented in the form $G={\bf \Omega}\cdot{\bf S}$ where ${\bf S}$ is the spin-one operator and $\Omega_z=0$. Thus the unitary operator $U(\vartheta,E=0)$ can be viewed as a sequence of infinitesimal rotations around a $\theta$-dependent axis ${\bf \Omega}$.  Since the sequence of rotations reduces to a single rotation around some axis ${\bf \bar{\Omega}}$, we find $U=\exp(i {\bf \bar{\Omega}}\cdot{\bf S})$ with $\bar{\Omega}_z=0$. It follows from the last equation that 
\begin{equation}
\label{b8}
{\bf Y}_{1,2}=\left(
\begin{array}{c}
\pm \frac{1}{2} e^{i\chi} \\ \frac{1}{\sqrt{2}} \\ \pm \frac{1}{2} e^{-i\chi}
\end{array}
\right) \;,\quad
{\bf Y}_3=\left(
\begin{array}{c}
\frac{1}{\sqrt{2}}e^{i\chi} \\ 0 \\ -\frac{1}{\sqrt{2}} e^{-i\chi}
\end{array}
\right) \;,
\end{equation}
where the phase $\chi=\chi(\vartheta;F)$ is a function of $\vartheta$ and $F$. The first two eigenvectors in Eq.~(\ref{b8}) correspond to complex-conjugated eigenvalues $\lambda_2^*(\vartheta) =\lambda_1(\vartheta)$ which determine the dispersive Wannier-Stark bands according to Eq.~(\ref{b7}). The third eigenvector corresponds to $\lambda_3=1$  which determines the flat Wannier-Stark bands.
We note that the proof of existence of the flat Wannier-Stark bands does not depend on the details of the dependence $\chi=\chi(\vartheta;F)$. Therefore flat Wannier-Stark bands exist for any rational orientation of the dc field. It is easy to generalize
this statement to any dc-biased chiral flat band in two dimensions with three bands, like the well-known 2d Lieb lattice \cite{Ramachandran:2017aa}.

\begin{figure}
\includegraphics[width=8.5cm]{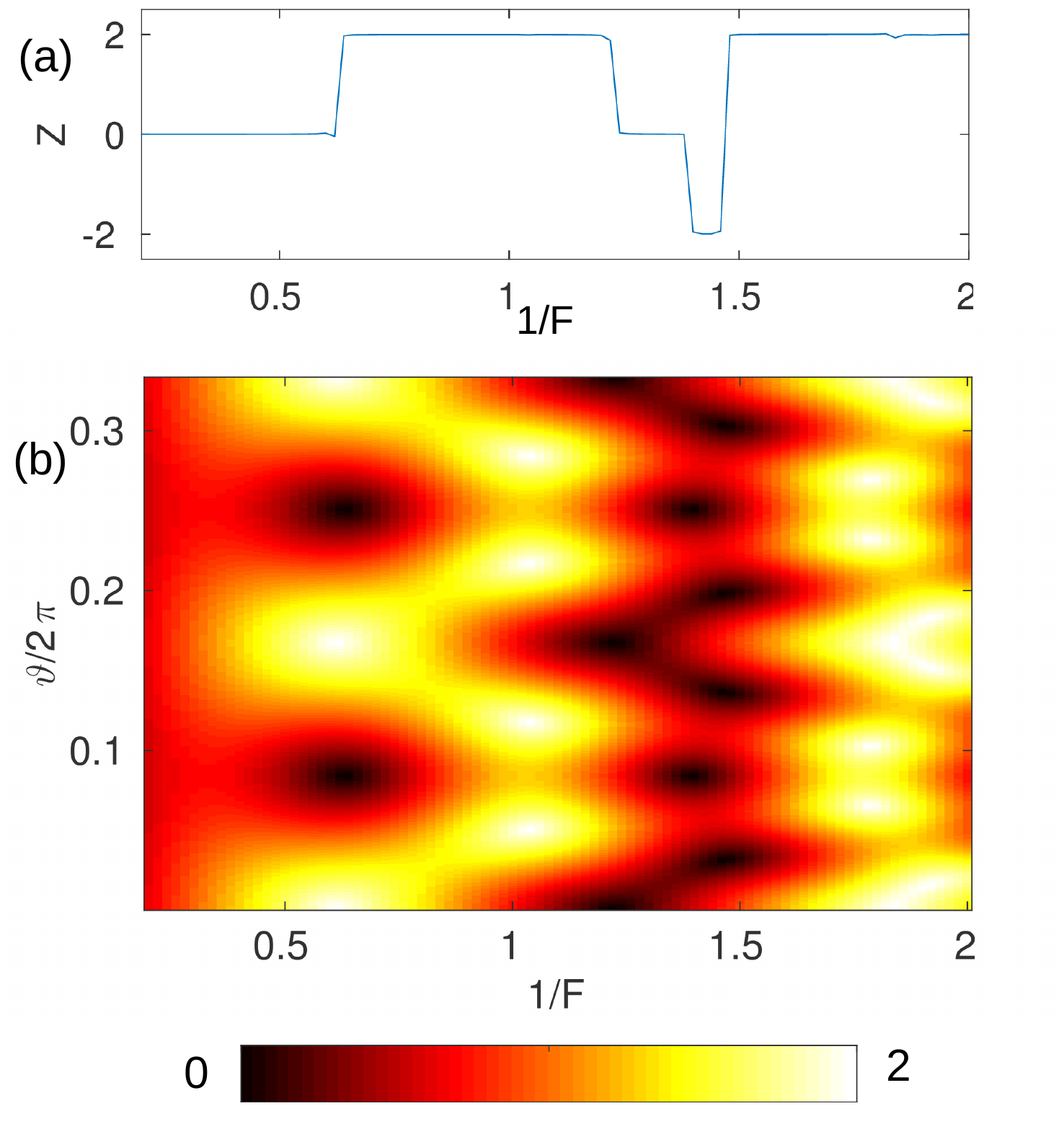}
\caption{(a) Winding number of the phase $\chi$ as the function of the inverse field magnitude. (b) The gap between the dispersive bands of the irreducible triplet as a function of $1/F$ and $\vartheta$. 
In this representation conical
intersections and vanishing gaps of the dispersive bands with the parent flat band appear as black spots.}
\label{fig4}
\end{figure}

The dispersive bands are sensitive to variations of $F$.  At the same time the bands have a topological invariant $Z$ -- the winding number of the relative phases of the eigenvector components:
\begin{equation}
\label{b9}
Z=\frac{1}{2\pi i} \int_0^{2\pi/3} e^{-i\chi}\frac{{\rm d}}{{\rm d} \vartheta} e^{i\chi} {\rm d}\vartheta 
\;,\quad Z=0,\pm 1,\ldots,
\end{equation}
which is closely related to the notion of Zak's phase \cite{PhysRevLett.62.2747,Atala:2013aa,PhysRevB.92.195144}.  The winding number ceases to be well defined whenever a degeneracy $\lambda_j(\vartheta)$ takes place. Thus, if we vary $F$, the quantity (\ref{b9})  may change its value at particular values of $F$ where the dispersive bands develop a conical intersection.  This is illustrated in Fig.~\ref{fig4}, where the Fig.\ref{fig4}(a) shows the winding number (\ref{b9}) as the function of $1/F$,  and Fig.\ref{fig4}(b) the gap between the
dispersive bands of the irreducible triplet
as a function of both inverse field magnitude and transverse quasimomentum. 
Conical intersections with vanishing gaps appear as black spots, the positions and number of which are seen to correlate with the
jump in value of the winding number $Z$ for the phase $\chi=\chi(\vartheta; F)$ in Eq.~(\ref{b8}). 

\begin{figure}
\includegraphics[width=8.5cm]{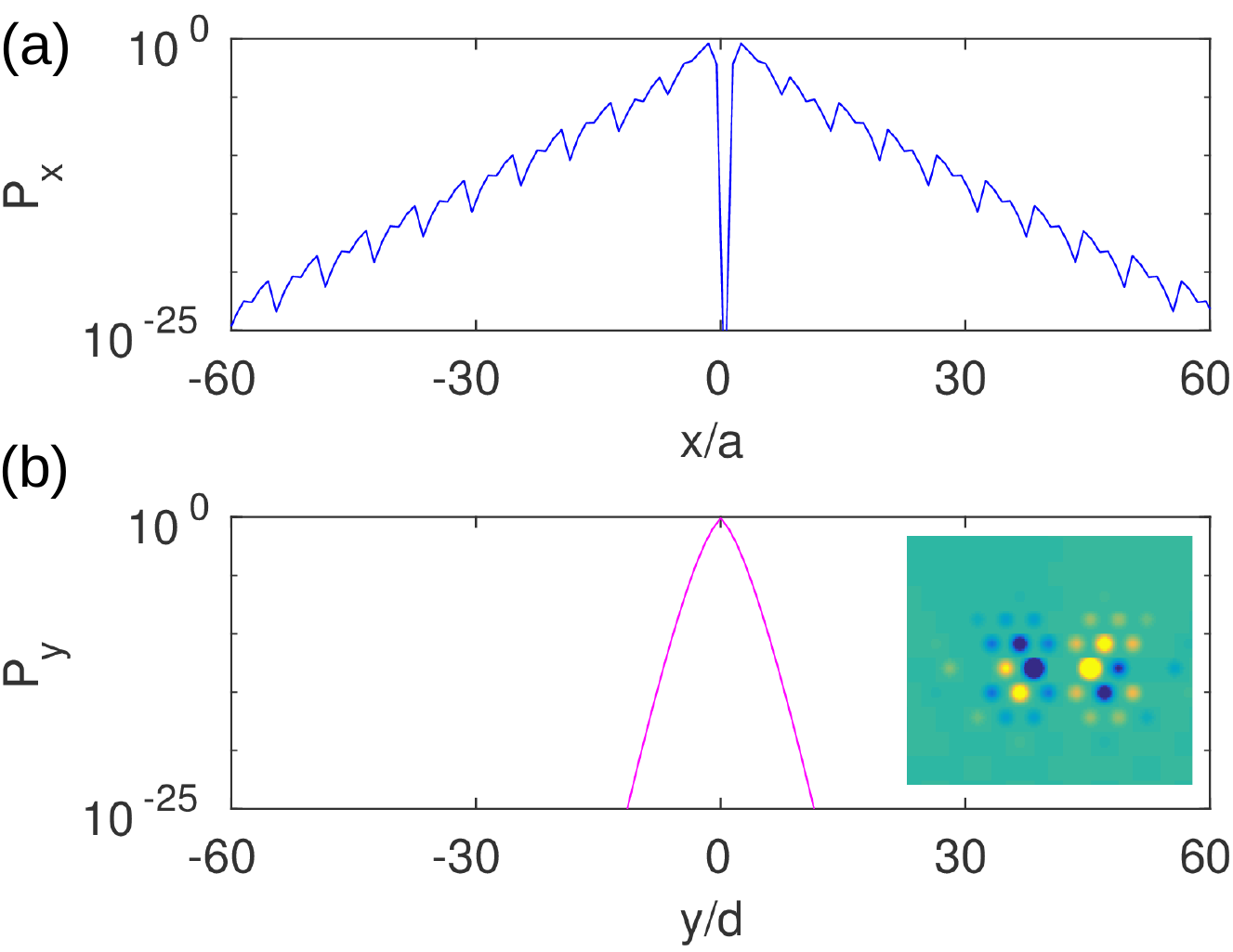}
\caption{ 
Integrated probabilities (a) $P(x)$ and (b) $P(y)$ in the logarithmic  scale for $F=4$. Inset:  The localized state in the tilted dice lattice for $F=4$. Occupation amplitudes $\Psi({\bf R}_j)$ of the lattice sites are shown as a color map with dark blue corresponding to $-1$ and bright yellow to $+1$.}
\label{fig5}
\end{figure}

Let us now discuss the particle eigenstates associated with flat bands.  Using the vector ${\bf Y}_3$ in Eq.~(\ref{b8}) as the initial condition and evolving it according to Eq.~(\ref{b3})  we obtain vector components as the function of two cyclic variables $\theta$ and $\vartheta$.  Next, taking the Fourier transform of the obtained functions over the variable $\theta$ we obtain site populations of the dual ladder system for a given value of the parameter $\vartheta$.  Analogously, taking the two-dimensional Fourier transform over the both cyclic variables we obtain site populations of the original 2D lattice, see inset of Fig.~\ref{fig5}.  It is interesting to compare the obtained localized state with CLS for $F=0$.  As it was already mentioned, the latter is given by an empty $C$ site surrounded by six $A$ and $B$ sites with alternating amplitudes $\pm1/\sqrt{6}$. It is seen in Fig.~\ref{fig5} that the center of gravity of the localized state in the tilted dice lattice is also an empty $C$ site.  However, the state itself  is {\sl not compact}. The results depicted in Fig.~\ref{fig5}, which shows integrated probabilities $P(x)=\int |\Psi({\bf R})|^2 {\rm d}y$ and $P(y)=\int |\Psi({\bf R})|^2 {\rm d}x$ on a logarithmic  scale, indicate that the state is exponentially localized in the direction orthogonal to ${\bf F}$ and super-exponentially in the direction parallel to ${\bf F}$. The absence of compact localization
for a flat band with short range hoppings is indicating nontrivial topology, as already observed through the winding number computations.

In conclusion, we showed that a biased dice lattice possesses flat Wannier-Stark bands which are supported by non-compact localized states. Formally, the existence  of flat bands follows from the algebraic  structure of the Floquet matrix $U$. However, one can also develop a `physical intuition' which tells that flat Wannier-Stark bands appear as soon as there are two alternative paths which recover the tunneling in the direction orthogonal to ${\bf F}$. For example, in the considered field orientation the tunneling between $B$ and $A$ sites is recovered through two $C$ sites, which are detuned from $A$ and $B$ sites by the Stark energy $\pm \sqrt{3}/2 F$. Similarly, for a field orientation parallel to the $x$ axis the tunneling between $C$ sites is recovered through $A$ and $B$ sites, which are detuned by  the energy $\pm F/2$. Clearly, the presence of two alternative paths requires the original 2D lattice to be a bipartite  lattice with majority and minority sublattices. We checked this conjecture for the Lieb lattice \cite{Ramachandran:2017aa} for two different field orientations $F_x/F_y=0$ and $F_x/F_y=1$. For these orientations the dual ladder system corresponds to biased stub and diamond chains, respectively. In both cases flat bands were found. Finally, we mention that non-compact localized states of the biased lattice do not converge to the compact localized states of the unbiased lattice as $F$ tends to zero but rather to a Bloch-like superposition of the compact  localized states. The details of the limit $F\rightarrow0$ will be discussed elsewhere.

\begin{acknowledgments}
AK acknowledges hospitality of IBS Center for Theoretical Physics of Complex Systems, where this work was conducted. This work was supported by the Institute for Basic Science, Project Code (IBS-R024-D1).
\end{acknowledgments}

\bibliography{flatband}

\end{document}